\documentclass[onecolumn,12pt]{article}
\usepackage{graphicx}
\usepackage{longtable}
\usepackage{lscape}
\usepackage{color}

\newcommand\simlt{\hspace{0.3em}\raisebox{0.4ex}{$<$}\hspace{-0.75em}\raisebox{-.7ex}{$\sim$}\hspace{0.3em}} 
\newcommand\simgt{\hspace{0.3em}\raisebox{0.4ex}{$>$}\hspace{-0.75em}\raisebox{-.7ex}{$\sim$}\hspace{0.3em}}

\begin{document}
\baselineskip 14pt

\title{Distortion of Magnetic Fields \\ 
in the Dense Core SL42 (CrA-E) \\
in the Corona Australis Molecular Cloud Complex
} 
\author{Ryo Kandori$^{1}$, Motohide Tamura$^{1,2,3}$, Masao Saito$^{2}$, Kohji Tomisaka$^{2}$, \\
Tomoaki Matsumoto$^4$, Ryo Tazaki$^{5}$, Tetsuya Nagata$^{6}$, Nobuhiko Kusakabe$^{1}$, \\
Yasushi Nakajima$^{7}$, Jungmi Kwon$^{3}$, Takahiro Nagayama$^{8}$, and Ken'ichi Tatematsu$^{2}$\\
{\small 1. Astrobiology Center of NINS, 2-21-1, Osawa, Mitaka, Tokyo 181-8588, Japan}\\
{\small 2. National Astronomical Observatory of Japan, 2-21-1 Osawa, Mitaka, Tokyo 181-8588, Japan}\\
{\small 3. Department of Astronomy, The University of Tokyo, 7-3-1, Hongo, Bunkyo-ku, Tokyo, 113-0033, Japan}\\
{\small 4. Faculty of Sustainability Studies, Hosei University, Fujimi, Chiyoda-ku, Tokyo 102-8160}\\
{\small 5. Astronomical Institute, Graduate School of Science, Tohoku University,}\\
{\small 6-3 Aramaki, Aoba-ku, Sendai 980-8578, Japan}\\
{\small 6. Kyoto University, Kitashirakawa-Oiwake-cho, Sakyo-ku, Kyoto 606-8502, Japan}\\
{\small 7. Hitotsubashi University, 2-1 Naka, Kunitachi, Tokyo 186-8601, Japan}\\
{\small 8. Kagoshima University, 1-21-35 Korimoto, Kagoshima 890-0065, Japan}\\
{\small e-mail: r.kandori@abc-nins.jp, r.kandori@gmail.com}}
\maketitle

\abstract{
Detailed magnetic field structure of the dense core SL42 (CrA-E) in the Corona Australis molecular cloud complex was investigated based on near-infrared polarimetric observations of background stars to measure dichroically polarized light produced by magnetically aligned dust grains. The magnetic fields in and around SL42 were mapped using 206 stars and curved magnetic fields were identified. On the basis of simple hourglass (parabolic) magnetic field modeling, the magnetic axis of the core on the plane of sky was estimated to be $40^{\circ} \pm 3^{\circ}$. The plane-of-sky magnetic field strength of SL42 was found to be $22.4 \pm 13.9$ $\mu$G. Taking into account the effects of thermal/turbulent pressure and the plane-of-sky magnetic field component, the critical mass of SL42 was obtained to be $M_{\rm cr} = 21.2 \pm 6.6$ M$_{\odot}$, which is close to the observed core mass of $M_{\rm core} \approx 20$ M$_{\odot}$. We thus conclude that SL42 is in a condition close to the critical state if the magnetic fields lie near the plane of the sky. Since there is a very low luminosity object (VeLLO) toward the center of SL42, it is unlikely this core is in a highly subcritical condition (i.e., magnetic inclination angle significantly deviated from the plane of sky). The core probably started to collapse from a nearly kinematically critical state. In addition to the hourglass magnetic field modeling, the Inoue \& Fukui (2013) mechanism may explain the origin of the curved magnetic fields in the SL42 region. 
}

\vspace*{0.3 cm}

\section{Introduction}
Studying magnetic fields associated with dense molecular cloud cores is important for revealing 1) the initial conditions of star formation, 2) the formation process of dense cores (i.e., structure formation in molecular clouds), and 3) the relationship between polarization and extinction (i.e., alignment of dust grains with respect to magnetic fields). These problems are not well understood, because the observations to probe the physical properties of dense cores are difficult, particularly observations of magnetic fields. The lack of magnetic information of dense cores leads to a consequent lack of important information for the understanding of star formation. 
\par
A popular method to measure the plane-of-sky magnetic field direction is the measurement of linearly polarized light produced by magnetically aligned dust grains in thermal emission (far-infrared to submm) or in the dichroic extinction of background starlight passing through dust grains (optical to near-infrared [NIR]). The Davis--Chandrasekhar--Fermi method (Davis 1951; Chandrasekhar \& Fermi 1953) is often employed to estimate the magnetic field strength of clouds or cores from these linear polarization data. 
\par
The most significant error factor in conducting magnetic field studies is the line-of-sight inclination angle of the magnetic axis of dense cores ($\gamma_{\rm mag}$). If a magnetic field experiences a distortion with an axisymmetric shape, e.g., hourglass-shaped fields (Kandori et al. 2017a, hereafter Paper I), we observe a significant depolarization pattern, particularly in the equatorial plane of the core, due to crossing of polarization vectors at the front and rear sides of the core. The shape of the depolarization pattern depends on the magnetic inclination angle. The angle $\gamma_{\rm mag}$ can be determined by comparing polarimetric observations with a three-dimensional (3D) model calculated for various inclination angles (Kandori et al. 2017b, hereafter Paper II; see also Kandori et al. 2020a, hereafter Paper VI). Note that few methods can measure $\gamma_{\rm mag}$ of dense cores. The depolarization pattern and magnetic inclination angle can be used to calibrate the observed polarization--extinction relationship (Kandori et al. 2018, hereafter Paper III; see also Paper VI). 
\par
Past studies that did not include $\gamma_{\rm mag}$ were unable provide details on the magnetic field strength for each object, because the ambiguity when the inclination angle is not known is significant. With knowledge of $\gamma_{\rm mag}$ based on the 3D analysis, we can discuss the universality and diversity of the magnetic field structure and total magnetic field strength of dense cores. 
To date, four low-mass dense cores with known 3D magnetic field structure and total strength (FeSt 1-457: Paper I, Barnard 68: Kandori et al. 2020b, Barnard 335: Kandori et al. 2020c, CB81: Kandori et al. 2020d) have been identified. 
The NIR polarimetric survey of low-mass dense cores and Bok globules is still in its infancy, and we need more objects for systematic/statistical studies. 
\par
As a part of our magnetic field survey of dense cores, we investigated the SL42 core (Hardegree-Ullman et al. 2013) in the Corona Australis molecular cloud (CrA) using NIR polarimetry. The SL42 core has many alternative designations, such as Cloud42/S42/SLDN42 (Sandqvist \& Lindroos 1976) and Core 5 (Yonekura et al. 1999). A clump including SL42 is named CrA-E in Bresnahan et al. (2018) based on large scale far-infrared (FIR) mapping with {\it Herschel}. In this paper, we use the name SL42 following Hardegree-Ullman et al. (2013). 
\par
The CrA complex exhibits an elongated \lq \lq head-tail'' morphology (e.g., Figure 1 of Bresnahan et al. 2018), having a dense \lq \lq head'' to the west and a diffuse \lq \lq tail'' to the east. Star-formation is taking place in the head region. The tail region consists of a \lq \lq north filament'' and \lq \lq south streamer'' (Bresnahan et al. 2018), and the SL42 core is located in the middle of the north filament. The distance to the CrA complex was thought to about 130 pc (Casey et al. 1998; Neuh\"{a}user \& Forbrich 2008). However, recent results using the {\it Gaia} data have provided a slightly more distant values of $154 \pm 4$ pc (Dzib et al. 2018) and $151 \pm 8$ pc (Zucker et al. 2019). In this paper, we will use 150 pc for the distance of the CrA complex. 
\par
The SL42 core was observed with the {\it Herschel} satellite and the 15-m Swedish ESO Submillimeter Telescope (SEST) (Hardegree-Ullman et al. 2013). The radius, mass, and central density of the core were $\approx 300''=0.23\ {\rm pc}$, $\approx 20$ M$_{\odot}$, and $\approx 1.0 \times 10^6$ cm$^{-3}$, respectively. Since Hardegree-Ullman et al. (2013) used 130 pc for the distance to SL42, we converted their physical quantities into values for a distance of 150 pc. The kinematic temperature of the core $T_{\rm k}$ was assumed to be 10 K. The density structure of the core was modeled using the Bonnor--Ebert sphere model (Ebert 1955; Bonnor 1956). The obtained best-fit Bonnor--Ebert parameter was $\xi_{\rm max} \approx 25$ (i.e., center to edge density contrast of $\approx 364$), which is far greater than the critical value of $\xi_{\rm max} = 6.5$, indicating that the core is unstable to gravitational collapse if there is no additional supporting force. The external pressure $P_{\rm ext}$ of the core was $(1.0 \times 10^6 / 364) \times T_{\rm k} = 2.8 \times 10^4$ K cm$^{-3}$ The center of the core determined using a column density map based on {\it Herschel} was (R.A., Decl.) = ($19^{\rm h}10^{\rm m}20.2^{\rm s}$, $-37^{\circ}08'26.0''$, J2000) (Hardegree-Ullman et al. 2013). The N$_2$H$^+$ ($J=1-0$) line width was about 0.52 km s$^{-1}$, which provides a turbulent velocity dispersion $\sigma_{\rm turb}$ of $0.21$ km s$^{-1}$. Note that we do not use C$^{18}$O ($J=2-1$) data observed with the SEST telescope, because significant CO depletion was observed in SL42 (Hardegree-Ullman et al. 2013). A very low luminosity object (VeLLO) candidate was found toward the center of SL42 in the {\it Herschel} 70 $\mu$m data (Bresnahan et al. 2018). 
Note that a VeLLO is defined as an young object with internal luminosity of $\le 0.1$ L$_{\odot}$ embedded in dense cloud cores (e.g., Young et al. 2004; Dunham et al. 2008; Kim et al. 2019). 
In addition, beyond the radius of SL42 ($7.4'$ from the center of the core), there is a weak-line T Tauri star H${\alpha}$16 at (R.A., Decl.) = ($19^{\rm h}06^{\rm m}23.8^{\rm s}$, $-37^{\circ}09'18.0''$, J1950) (Marraco \& Rydgren 1981; Batalha et al. 1998; Gregorio-Hetem \& Hetem 2002). 
\par
In this study, wide-field background star polarimetry at NIR wavelengths was conducted for SL42. The plane-of-sky magnetic field structure was revealed using stars in and around the core radius. The total magnetic field strength of the core was estimated based on the Davis--Chandrasekhar--Fermi method (Davis 1951; Chandrasekhar \& Fermi 1953) and 3D magnetic field modeling of the core. Using the resulting magnetic field information, the kinematical stability and the formation scenario of SL42 is discussed. 

\section{Observations}
The NIR polarimetric observations of the SL42 core region in the CrA cloud complex was conducted using the IRSF 1.4-m telescope and $JHK_s$ simultaneous polarimeter SIRPOL (Kandori et al. 2006, see also Nagayama et al. 2003 for camera module). IRSF/SIRPOL provides large field of view ($7.7' \times 7.7'$ with a scale of $0.45$ $''/{\rm pixel}$), which enable us to cover nearby dark cloud cores with a single telescope pointing. SIRPOL is a single-beam polarimeter, which consists of a rotating half-wave plate and a wire-grid polarizer. 
\par
The fluctuations of measured polarization degree during exposures are typicaly $\approx 0.3\%$. The instrumental polarization over the field of view is confirmed to be less than $0.3\%$. The uncertainty of the zero point angle of the polarimeter is less than $3\%$ (Kandori et al. 2006; Kusune et al. 2015). A polarized standard star RCrA\#88 was observed on July 13, 2017 to obtain $P_H = 2.82\% \pm 0.09\%$ and $\theta_H = 91.9^{\circ} \pm 0.9^{\circ}$, which are consistent with data in the literature ($P_H = 2.73\% \pm 0.07\%$, $\theta_H = 92^{\circ} \pm 1^{\circ}$, Whittet et al. 1992). 
\par
Observations of SL42 (mosaic of ten images) were conducted on the nights of June 14, 21, 23, 25, and 29, and July 3, 4, 6, 7, and 11 in 2017. Exposures of 15-s were performed at four half-wave plate angles, in the sequence of $0^{\circ}$, $45^{\circ}$, $22.5^{\circ}$, and $67.5^{\circ}$, at ten dithered positions (one set). Total exposure time was 1500-s (10 sets) for each half-wave plate angle. Typical seeing conditions were $\approx 1.4''$ ($\approx 3$ pixels) at $H$. 
\par
The acquired data were reduced using the Interactive Data Language (IDL). Following the methods described in Kandori et al. (2007), we performed dark subtraction, flat correction, frame combine after registration. We obtained a combined image of Stokes $I$ and four images combined with respect to the half-wave plate angle of $I_{0^{\circ}}$, $I_{45^{\circ}}$, $I_{22.5^{\circ}}$, and $I_{67.5^{\circ}}$. Twilight flat was used to calibrate these observations, and we confirmed that the process did not produce false signals or instrumental polarizations (Kandori et al. 2020d). 
\par
Point sources with a peak greater than $10\sigma$ from local sky background were cataloged on the Stokes $I$ image. Using the cataloged positions of stars, aperture polarimetry was performed on the $I_{0^{\circ}}$, $I_{45^{\circ}}$, $I_{22.5^{\circ}}$, and $I_{67.5^{\circ}}$ images. We set aperture radius to the same as the full width at half maximum (FWHM) of each mosaic image, and sky radius and width of sky annulus were set to 10 and 5 pixels, respectively. Relatively small aperture size was used in order to avoid stellar flux contamination in a crowded field. We did not employ the point spread function (psf) fitting photometry, because relative photometry is important in polarimetry. The goodness of fits can vary on each half-wave plate angle image, and this causes the systematic error in polarimetric measurements. The stars with the photometric uncertainty of greater than 0.1 mag were removed from the list. 4319 sources were detected in the $H$ band. The limiting magnitudes were 18.0 mag in the $H$ band. 
\par
The Stokes parameters for each star were derived from the equations $I = (I_{0}+I_{45}+I_{22.5}+I_{67.5})/2$, $Q = I_{0} - I_{45}$, and $U = I_{22.5} - I_{67.5}$. Polarization degree $P$ and angle $\theta$ were obtained with $P = \sqrt{Q^2 + U^2}/I$ and $\theta = 0.5 {\rm atan} (U/Q)$. $P$ tends to be overestimated particularly for low $S/N$ data. To correct this effect, we used $P_{\rm db} = \sqrt{P^2 - \delta P^2}$ (Wardle \& Kronberg 1974). We use debiased $P$ for the discussion of polarization in the followings. 

In the present study, we discuss the results obtained in the $H$ band, for which dust extinction effects are less severe than in the $J$ band, and the polarization efficiency is greater than in the $K_s$ band.

\section{Results and Discussion}
\subsection{Distortion of Magnetic Fields}
Figure 1 presents the finding chart of our observations. The background image is the column density map based on the {\it Herschel} data (Andr\'{e} et al. 2010; Bresnahan et al. 2018). The SL42 region is located near the center of the image, and our IRSF/SIRPOL observation region is enclosed by the white line. Figure 2 shows the observed polarization vectors (yellow lines) on the Stokes $I$ mosaic image in the $H$ band. Polarization vectors generally flow from north-east to south-west. The most significant feature on the vector map is the curved structure in the polarization vectors, which is particularly prominent in the center to south part. There is a slightly curved structure in the northernmost part of the mosaic image. Figure 3 is the same as Figure 2 but the background image is the {\it Herschel}-based column density map and the white circle shows the radius (300$''$) of the SL42 core (Hardegree-Ullman et al. 2013). 
\par
There are at least two possible interpretations of curved magnetic field structure. At a glance, the southern curved field structure seems to be associated with the SL42 core. The magnetic field lines appear to wrap around the core. This resembles a mechanism proposed by Inoue \& Fukui (2013) in which the interaction between a shock wave and the core can create a bending magnetic field structure. 
Another possibility is the existence of hourglass-shaped fields with a magnetic center offset from the center of the mass distribution. We observed a similar geometry in the CB81 core in the Pipe Nebula (Kandori et al. 2020b). In this case, a nonuniform initial density or magnetic field distribution can be compressed by turbulence or shocks to create the \lq \lq offset hourglass field structure''. In this study, we first model the data using the offset hourglass field both in 2D and 3D. Then we will discuss the possibility of the Inoue \& Fukui (2013) mechanism in Section 3.4. 
\par
Figure 4 shows schematic figures for these two scenarios. In the figure, the white lines show the magnetic field line, and the red plus signs show the center of the core. The background image shows the column density distribution. In panels (a) and (b) of Figure 4, the magnetic field lines darken toward the lower-left corner. In panel (a), this corresponds to the small polarization degree due to a depolarization effect prominent in the steep field curvature part. In panel (b), this corresponds to a decrease of the winding magnetic fields because a shock wave arrives from the direction of the upper-right corner and the direction of the lower-left corner is behind the core. 
\par
In each panel of Figure 4, we used the function $y=g+gCx^2$ to draw the magnetic field lines. In the function, $g$ specifies each magnetic field line and $C$ determines the steepness of the curvature of the parabolic function. Since $g$ is included in the second term, the steepness of curvature increases with distance from the $x$-axis. For panel (a), the feature is consistent with the analytically described hourglass-shaped magnetic field model (Mestel 1966; Ewertowski \& Basu 2013; Myers et al. 2018). For panel (b), the function $y=g+Cx^2$ may provide a better description of the wrapping of magnetic fields around the edge-on cylinder. However, the present study, as well as the study by Hardegree-Ullman et al. (2013), assumes that the SL42 core is spherical in shape. In this case, wrapping of the magnetic fields also occurs at the front and rear of the core, which results in a field component with a polarization direction parallel to the direction of the shock wave. The parallel field is superimposed on the wrapping magnetic fields of the core, so that the curvature of the magnetic field lines located at the rear of the core against the shock wave can be steeper than $y=g+Cx^2$. Thus, for describing this configuration, we concluded that the function $y=g+gCx^2$ is better than $y=g+Cx^2$. Note that the function to model the Inoue \& Fukui (2013) mechanism depends on both the density structure of the blob and the viewing angle (e.g., see Tahani et al. 2019; Reissl et al. 2018). The function form presented here is just a simple representation of the bending field shape predicted by Inoue \& Fukui (2013). 
\par
On the basis of the discussion above, we obtained similar curved magnetic field geometries for (a) offset hourglass and (b) Inoue--Fukui mechanism as shown in Figure 4. The south-east part in the field geometry is the same for (a) and (b). 
If the north-west field component is prominent, we can conclude the scenario (a) is plausible, because the scenario (b) cannot create such a structure. If not, it is not easy to discriminate them. In this case, there are two possible scenarios. We discuss two scenarios in Section 3.4 to assess the origin of the curved magnetic fields. 
\par
Figure 5 shows the $P_H$ versus $H-K_s$ (i.e., $A_V$) relationship. To estimate $A_V$, we assumed a region of declination of $\ge$ $-36^{\circ}56'$ as an extinction-free reference field, and obtained 0.155 mag as the average $H-K_s$ color of stars in the reference field. The value is close to the average $H-K_s$ color of typical field stars (0.15 mag, Lada et al. 1994). All the stars with $H-K_s \ge 0.7$ mag are located within the radius of SL42. A linear fitting to these stars resulted in a slope of $2.71 \pm 0.31$ \% mag$^{-1}$ (dashed line in Figure 5). If we include all the source in the fitting, the resulting slope is $13.56 \pm 0.63$ \% mag$^{-1}$ (dotted line), which is close to the upper limit of the polarization efficiency for the interstellar medium ($\approx 14$, Jones 1989). Note that the distribution of all the stars is distorted and it may not be good to fit them using linear fitting which assumes homoscedastic Gaussian scatter. Therefore, the dotted line in Figure 5 is just a reference. 
%
In Figure 5, there are several stars with high $P_H$ ($\simgt 5$ \%) and without large $A_V$ ($\simlt 5$ mag). Though these sources are included in the 2D and 3D magnetic field analysis, they do not affect the conclusion, because their number is small compared with the whole sample ($N = 206$). 


\subsection{Parabolic Model}
Figure 6 shows the configuration of the magnetic field (solid white lines) estimated using a parabolic function and its shift and rotation. 
%
In the figure, 206 polarization vectors having $P_H \ge 0.5$ \% and $P_H/\delta P_H \ge 5$ are included. For the parabolic function, $y=g+gCx^2$ was used, where $g$ specifies the magnetic field lines and $C$ determines the degree of curvature. $\theta_{\rm mag}$ is the position angle of the magnetic field direction (from north through east). Note that we used the parabolic function in the $90^{\circ}$-rotated form so that, when $\theta_{\rm mag}$ is $0^{\circ}$, the direction corresponds to the direction of declination. The best-fit parameters were determined to be $\theta_{\rm mag} = 40^{\circ} \pm 3^{\circ}$ and $C = 5.0(\pm 0.3) \times 10^{-6}$ arcsec$^{-2}$. 
\par
A parabolic function was employed because this is the simplest form of approximating the analytically described hourglass-shaped magnetic field model (Mestel 1966; Ewertowski \& Basu 2013; Myers et al. 2018; see also Paper VI for a comparison of the parabolic function to the hourglass model). Note that the bending field structure in the south-east part of the figure resembles the model by Inoue \& Fukui (2013). 
\par
In the fitting procedure, the observational error associated with each star was taken into account when calculating 
\begin{equation}
\chi^2 = \sum_{i=1}^n \frac{(\theta_{\rm obs,{\it i}} - \theta_{\rm model}(x_i,y_i))^2}{\delta \theta_i^2}, 
\end{equation}
where $n$ is the number of stars, $x$ and $y$ are the coordinates of these stars, $\theta_{\rm obs}$ and $\theta_{\rm model}$ denote the polarization angles from observations and from the model, and $\delta \theta_i$ is the observational error. 
The coordinate origin of the parabolic function is R.A.=19$^{\rm h}$09$^{\rm m}$42$.\hspace{-3pt}^{\rm s}$46, Decl.=-36$^{\circ}$59$'$26$.\hspace{-3pt}''5$ (J2000), determined by searching for the minimum $\chi^2$ for the variables $x$ and $y$ in and around the core. The center coordinate of the core (the peak of the column density distribution based on {\it Herschel}) is R.A.=19$^{\rm h}$10$^{\rm m}$20$.\hspace{-3pt}^{\rm s}$2, Decl.=-37$^{\circ}$08$'$26$.\hspace{-3pt}''0$ (J2000). The angular distance between these two centers is $11.72'$, which roughly corresponds to one core diameter ($10'$). 
\par
The parabolic fitting was satisfactory, since the standard deviation of the residual angles, $\theta_{\rm res} = \theta_{\rm obs} - \theta_{\rm fit}$, was smaller when using the parabolic function ($\delta \theta_{\rm res} = 28.9^{\circ}$) than in the case of a uniform field ($\delta \theta_{\rm res} = 55.0^{\circ}$). However, a closer look at the data reveals that there are areas where the model and observations deviate systematically. This leads to an overestimate of the standard deviation of the residual angle. Thus, we divided the stars using the coordinate grid with $300''$ width in the R.A. and Decl. direction, and calculated the mean $\theta_{\rm res}$ in each grid box. The result is shown in Figure 7. The values were treated as an offset deviation angle in each box, and subtracted from the $\theta_{\rm res}$ values of stars falling in each box. We found $17.43^{\circ}$ for the offset-subtracted $\delta \theta_{\rm res}$ value. Since the grid size is the same as the radius of the core, we believe that we are subtracting a sufficiently large structure in this analysis. Figure 8 shows the histogram of the offset subtracted $\theta_{\rm res}$. Employing $600''$ (core diameter) and $150''$ (1/2 core radius) for grid size, we obtained $25.23^{\circ}$ and $11.66^{\circ}$ for $\theta_{\rm res}$. We thus used $8^{\circ}$ as the uncertainty of the offset-subtracted $\delta \theta_{\rm res}$. The intrinsic dispersion, $\delta \theta_{\rm int} = (\delta \theta_{\rm res}^2 - \delta \theta_{\rm err}^2)^{1/2}$, estimated using the parabolic fitting, was found to be $17.15^{\circ}$ (0.299 radian), where $\delta \theta_{\rm err}$ is the standard deviation of the observational error in the polarization measurements. 
\par
The polarization vectors used to obtain the above polarization angle dispersion values include the vectors located outside of core's boundary. We thus check the angle dispersion using the stars fallen inside the core radius (25 stars), and obtained the offset subtracted $\theta_{\rm res}$ as $12.58^{\circ}$, $16.28^{\circ}$, and $18.87^{\circ}$, for the grid width of $150''$, $300''$, and $600''$, respectively. Therefore, the value of $\theta_{\rm res} = 17.43^{\circ} \pm 8.00^{\circ}$ employed above includes the polarization angle dispersion for the core. Note that in the angle dispersion calculation, outliers with more than $45^{\circ}$ angle deviation (three out of 25 vectors) were rejected. 
\par
Assuming frozen-in magnetic fields, the intrinsic dispersion of the magnetic field direction, $\delta \theta_{\rm int}$, can be attributed to the perturbation of the Alfv\'{e}n wave by turbulence. The strength of the plane-of-sky magnetic field ($B_{\rm pos}$) can be estimated from the relationship $B_{\rm pos} = C_{\rm corr} (4 \pi \rho)^{1/2} \sigma_{\rm turb} / \delta \theta_{\rm int}$, where $\rho$ and $\sigma_{\rm turb}$ are the mean density of the core and the turbulent velocity dispersion (Chandrasekhar \& Fermi 1953), and $C_{\rm corr} = 0.5$ is a correction factor suggested by theoretical studies (Ostriker et al. 2001; see also, Padoan et al. 2001; Heitsch et al. 2001; Heitsch 2005; Matsumoto et al. 2006). 
%
%
Using the mean density ($\rho = 3.11 \times 10^{-20}$ g cm$^{-3}$) and turbulent velocity dispersion ($\sigma_{\rm turb} = 0.21$ km s$^{-1}$) from Hardegree-Ullman et al. (2013) and $\delta \theta_{\rm int}$ derived in this study, we obtained $B_{\rm pos} = 22.4$ $\mu$G. For error estimation, we assumed that the turbulent velocity dispersion and mean density values were accurate within 30\%. Thus, the uncertainty in $B_{\rm pos}$ was estimated to be 13.9 $\mu$G. Note that the uncertainty of $8^{\circ}$ of $\delta \theta_{\rm res}$ is included in the error calculation..

\subsection{3D Magnetic Field}
The 3D magnetic field modeling was performed following the same procedure described in a previous paper (Section 3.1 of Paper VI). The 3D version of the simple parabolic function, $z(r,\varphi,g) = g + gCr^2$ in the cylindrical coordinate system $(r,z,\varphi)$, was used for modeling the core magnetic fields, where $g$ specifies the magnetic field lines, $C$ is the curvature of the lines, and $\varphi$ is the azimuth angle (measured in the plane perpendicular to $r$). Using this function, the magnetic field lines are axially symmetric around the $r$ axis. 
\par
In the model, the amount of polarization per unit volume was assumed to be proportional to the mass in the volume. For the density structure of the core, we employed the Bonnor--Ebert sphere with the solution parameter of $\xi_{\rm max} = 25$ (Hardegree-Ullman et al. 2013). The orientation of polarization generated by dust grains in each cell was assumed to be parallel to the orientation of magnetic field at the location of the cell. 
\par
The 3D model was virtually observed after rotating in the line of sight ($\gamma_{\rm mag}$) and the plane-of-sky ($\theta_{\rm mag}$) directions. $\gamma_{\rm mag}$ is the line-of-sight inclination angle measured from the plane-of-sky. In this procedure, the Stokes parameters in each cell of the model core were integrated toward the line of sight. The analysis was performed in the same manner described in Section 3.1 of Paper VI. The resulting polarization vector maps for the 3D parabolic model for various $\gamma_{\rm mag}$ are shown in Figure S1 of Paper VI. 
%
%
\par
Since the polarization distributions in the model core differ from one another depending on the viewing angle toward the line of sight, $\chi^2$ fitting of these distributions with the observational data can be used to restrict the line-of-sight magnetic inclination angle and 3D magnetic curvature. 
\par
Figure 9 summarizes the distribution of $\chi^2_\theta$ calculated using the model and observed polarization angles as 
\begin{equation}
\chi^2_\theta = \sum_{i=1}^n \frac{(\theta_{\rm obs,{\it i}} - \theta_{\rm model}(x_i,y_i))^2}{\delta \theta_i^2}, 
\end{equation}
where $n$ is the number of stars, $x$ and $y$ show the coordinates of stars, $\theta_{\rm obs}$ and $\theta_{\rm model}$ denote the polarization angle from observations and the model, and $\delta \theta_i$ is the observational error. The optimal magnetic curvature parameter, $C$, was determined at each inclination angle $\gamma_{\rm mag}$ to obtain $\chi^2_\theta$. 
\par
From the polarization angle fitting, it is clear that a small inclination angle ($\gamma_{\rm mag} \simlt 50^{\circ}$) and pole-on geometry ($\gamma_{\rm mag} = 90^{\circ}$) are unlikely. The distribution of $\chi_{\theta}^2$ is relatively flat for the region $60^{\circ} \simgt \gamma_{\rm mag} \simgt 85^{\circ}$, and there is a minimization point at $\gamma_{\rm mag} = 80^{\circ}$. Though the estimated 1 sigma uncertainty is $8^{\circ}$, we conclude that $\gamma_{\rm mag}$ is $75^{\circ}$ with an uncertainty of $15^{\circ}$. The best magnetic curvature obtained at $\gamma_{\rm mag} = 75^{\circ}$ was $2.87 \times 10^{-6}$ arcsec$^{-2}$. 
\par
It is notable that the $\chi^2$ values in Figure 9 are very large (about $6 \times 10^4$). The value is still large if it is divided by the number of stars ($N=206$). We tried to fit the large-scale hourglass-shaped structure, and in addition to the observational error, each data point has a scatter from Alfv\'{e}n waves. In the equation (2), we included the observational error in the denominator of the $\chi^2$ equation, but we could not include the scatter caused by Alfv\'{e}n waves. That's why we obtained large $\chi^2$ values. In Section 3.2, we obtained the intrinsic scatter caused by Alfv\'{e}n waves as $\delta \theta_{\rm int} = 17.15^{\circ}$. If we include the square of the value in the denominator of the Equation (2), and further divide Equation (2) by the number of stars ($N=206$), we obtained the reduced $\chi^2 = 2.21$ at the minimization point. 
%
\par
Figure 10 shows the best-fit 3D parabolic model along with the observed polarization vectors. The background image is the column density map of the model core processed using the line integral convolution (LIC) technique (Cabral \& Leedom 1993). We used the publicly available IDL code developed by Diego Falceta Gon\c{c}alves. The direction of the LIC \lq \lq texture'' is parallel to the magnetic field direction. The direction of the model vectors generally agrees with the observations. The standard deviation of the differences in the plane-of-sky polarization angles between the 3D model and the observations is $15.32^{\circ}$, which is close to the offset (measured with $300''$ grid) subtracted $\delta \theta_{\rm res}$ value. 

\subsection{Magnetic Properties of the Core and the Origin of the Curved Magnetic Fields}
Here, we estimate and discuss the core's magnetic properties, first without 3D correction (plane-of-sky component) and then considering the validity of 3D correction based on the inclination angle obtained using the offset hourglass modeling. 
\par
The plane-of-sky magnetic field strength averaged for the whole core was determined to be $22.4 \pm 13.9$ $\mu$G in Section 3.2. The magnetic support of the core against gravity can be investigated using the parameter $\lambda = (M/\Phi)_{\rm obs} / (M/\Phi)_{\rm critical}$, which represents the ratio of the observed mass-to-magnetic flux ratio to a critical value, $(2\pi G^{1/2})^{-1}$, suggested by theory (Mestel \& Spitzer 1956; Nakano \& Nakamura 1978). Based on the plane-of-sky magnetic field strength, we determined a value of $\lambda = 2.0 \pm 0.7$ (i.e., magnetically supercritical). The magnetic critical mass of the core of $9.9 \pm 4.7$ M$_{\odot}$ was lower than the observed core mass of $M_{\rm core} \approx 20$ M$_{\odot}$. Note that this does not necessarily imply the gravitational collapse of the core, because there are additional thermal/turbulent pressure components. 
\par
The critical mass of SL42, taking into account both magnetic and thermal/turbulent support effects is $M_{\rm cr} \simeq M_{\rm mag} + M_{\rm BE}$ (Mouschovias \& Spitzer 1976; Tomisaka, Ikeuchi \& Nakamura 1988; McKee 1989) and was determined to be $9.9 + 11.3 = 21.2 \pm 6.6$ M$_{\odot}$, where $M_{\rm BE} = 11.3 \pm 4.7$ M$_{\odot}$ is the Bonnor--Ebert mass calculated using an assumed kinematic temperature of 10 K , a turbulent velocity dispersion of 0.21 km s$^{-1}$ (equivalent to a temperature of 13 K, calculated from Hardegree-Ullman et al. 2013), and an external pressure of $2.86 \times 10^4$ K cm$^{-3}$ (calculated from Hardegree-Ullman et al. 2013). The estimated $M_{\rm cr}$ value is comparable to the observed core mass $M_{\rm core}$, indicating that SL42 is in a nearly critical state if the direction of the magnetic fields lies in the plane of sky. 
\par
The relative importance of magnetic fields with respect to the support of the core was investigated using the ratios of the thermal and turbulent energy to the magnetic energy, $\beta \equiv 3C_{\rm s}^2/V_{\rm A}^2$ and $\beta_{\rm turb} \equiv \sigma_{\rm turb, 3D}^2 / V_{\rm A}^2 = 3\sigma_{\rm turb, 1D}^2 / V_{\rm A}^2$, where $C_{\rm s}$, $\sigma_{\rm turb}$, and $V_A$ denote the isothermal sound speed at 10 K, the turbulent velocity dispersion, and the Alfv\'{e}n velocity. These ratios were found to be $\beta = 0.83 \pm 0.56$ and $\beta_{\rm turb} = 1.08 \pm 0.97$. Though uncertainties are large, the thermal, turbulent, and magnetic energies are consistent with being in equipartition, if the direction of magnetic fields is in the plane of sky.
\par
As noted in the Introduction, a VeLLO candidate was found toward the center of SL42 (Bresnahan et al. 2018). In addition, there is a weak-line T Tauri star H${\alpha}$16 (Marraco \& Rydgren 1981; Batalha et al. 1998; Gregorio-Hetem \& Hetem 2002) in the vicinity of the core. It is thus reasonable to treat SL42 as a protostellar core. SL42 is in a nearly kinematically critical state with the plane-of-sky magnetic field component. If the magnetic inclination angle of SL42 deviates from the plane of sky, the physical status of SL42 should be subcritical, which is not consistent with it being a protostar. Thus, from the magnetic properties of the core and the existence of young star(s), it is most likely that the SL42 core is associated with nearly plane-of-sky magnetic fields, and the core has started a collapse from a nearly kinematically critical state. If this is true, the offset hourglass field scenario of SL42, which provides a large inclination angle of $\gamma_{\rm mag} = 75^{\circ}$ in its 3D analysis in Section 3.3, is probably inaccurate. 
\par
Here, we derive the magnetic properties of SL42 with a 3D correction based on the hourglass modeling. The inclination correction factor is $1/\cos{75^{\circ}}=3.86$. The total magnetic field strength averaged for the whole core is $B_{\rm tot} = 86.45$ $\mu$G, which is very strong for a mean density of $3.11 \times 10^{-20}$ g cm$^{-3}$ ($7.99 \times 10^3$ cm$^{-3}$). The ratio of the mass-to-magnetic flux ratio to the critical value $\lambda$ is $0.52$, suggesting that SL42 is in a magnetically subcritical state. The critical mass of the core is $M_{\rm cr} \simeq M_{\rm mag} + M_{\rm BE} = 49.4$ M$_{\odot}$, where $M_{\rm mag} = 38.1$ M$_{\odot}$ and $M_{\rm BE} = 11.3$ M$_{\odot}$, indicating a subcritical state for SL42. For energy comparison, $\beta = 0.06 \pm 0.03$ and $\beta_{\rm turb} = 0.07 \pm 0.06$. 
%
%
Based on these results, the magnetic field is extremely strong and maintains the core as being subcritical, and for energy equipartition, the extremely strong magnetic fields disbalance the distribution of the energy. Though it is possible that SL42 is a peculiar core, these quantities do not fit with the existence of young stars in and around SL42. 
\par
%
%
For the scenario of a curved field in SL42, another possibility is the Inoue \& Fukui (2013) mechanism. 
Figure 11 shows the magnetic field distribution for scenario (b). As discussed in Section 3.1, the use of the function $y = g + gCx^2$ is reasonable to describe the plane-of-sky polarization vector distribution of SL42.
%
The expected direction of the shock wave propagation is $\theta_{\rm shock} \approx 130^{\circ}$, which is $90^{\circ}$ from $\theta_{\rm mag}$. The $\theta_{\rm shock}$ direction is roughly toward the Upper Centaurus Lupus OB association. The influence of the OB association on the CrA complex is shown and discussed in Bresnahan et al. (2018) and Harju et al. (1993). For scenario (b), the northern slightly curved magnetic component can be ignored in the fitting. 
\par
Figure 12 shows the global magnetic field structure ($8^{\circ} \times 8^{\circ}$) around the CrA complex based on 353-GHz dust polarization data taken by the {\it Planck} satellite. The box enclosed by the white line shows the surveyed region with the NIR polarization. Complex magnetic field structures are observed toward the CrA complex, although the resolution of the image is relatively large ($5'$). The direction of the magnetic fields is roughly east--west, which is roughly parallel to the orientation of the cloud complex (Planck Collaboration XXXV 2016). 
A bending structure is apparent in our surveyed region (white line), which can also be observed in our NIR polarimetry (see the center to the south region in Figure 2). From the {\it Planck} data, it is evident that there is a single curved structure in our surveyed region. Thus, scenario (b) in Figure 4 may be more likely than scenario (a) based on the morphology of large scale magnetic fields. 
\par
There are at least two scenarios for the explanation of the magnetic field distribution in and around the SL42 core. The scenario (a) is the offset hourglass structure, and the scenario (b) is the Inoue \& Fukui (2013) mechanism. 
\par
The 3D analysis of the offset hourglass-shaped magnetic field structure (Section 3.2) resulted in a very subcritical condition. This result cannot explain the existence of young stars possibly associated with the SL42 core. Though the SL42 core might be a peculiar core, this is a weak point of the scenario (a). 
\par
The 3D modeling of the Inoue--Fukui mechanism is not easy. In the model, the shock wave interacts with a dense core, and magnetic field lines at the surface of the core can change direction to wrap around the core. The white lines in Figure 11 are only rough approximation that take into account the characteristics of the model. The comparison of the observations with theoretical simulations is beyond the scope of this paper. The incompleteness in the modeling based on the Inoue--Fukui mechanism is a weak point of the scenario (b). 
\par
As discussed above, the choice of the line-of-sight magnetic inclination angle largely deviated from the plane of sky can lead to the highly subcritical condition of the SL42 core. This does not fit in the existence of young stars in and around SL42. We thus expect a nearly plane-of-sky magnetic field geometry for SL42, which can result in a nearly critical state. 

\section{Summary and Conclusion}
Detailed magnetic field structure of the dense core SL42 (CrA-E) in the Corona Australis molecular cloud complex was investigated based on NIR polarimetric observations of background stars to measure dichroically polarized light produced by magnetically aligned dust grains. The magnetic fields in and around SL42 were mapped using 206 stars and curved magnetic fields were identified. Based on simple hourglass (parabolic) magnetic field modeling, the magnetic axis of the core on the plane-of-sky was estimated to be $\theta_{\rm mag} = 40^{\circ} \pm 3^{\circ}$. The plane of sky magnetic field strength of SL42 was found to be $22.4 \pm 13.9$ $\mu$G. Taking into account the effects of thermal/turbulent pressure and the plane-of-sky magnetic field component, the critical mass of SL42 is found to be $M_{\rm cr} = M_{\rm mag} + M_{\rm BE} = 21.2 \pm 6.6$ M$_{\odot}$, which is close to the observed core mass of $M_{\rm core} \approx 20$ M$_{\odot}$. In the equation, the magnetic critical mass is $M_{\rm mag} = 9.9 \pm 4.7$ M$_{\odot}$ (magnetically supercritical) and the Bonnor--Ebert mass is $M_{\rm BE} = 11.3 \pm 4.7$ M$_{\odot}$. We conclude that SL42 is in a condition close to the critical state if the magnetic fields lie near the plane of the sky. Since there is a very low luminosity object (VeLLO) toward the center of SL42, it is unlikely the core is in a highly subcritical condition (i.e., magnetic inclination angle significantly deviated from the plane of sky). The core probably started to collapse from a nearly kinematically critical state. 
In addition to the hourglass magnetic field modeling, the Inoue \& Fukui (2013) mechanism may explain the origin of the curved magnetic fields in the SL42 region. The curved magnetic field structure could be created by a shock wave interacting with a dense blob with magnetic fields subsequently swept by the shock wrapping around the blob. 


\subsection*{Acknowledgement}
The contributions of Tsuyoshi Inoue are gratefully acknowledged. We are also grateful to the staff of SAAO for their help during the observations. We wish to thank Tetsuo Nishino, Chie Nagashima, and Noboru Ebizuka for their support in the development of SIRPOL, its calibration, and stable operation with the IRSF telescope. The IRSF/SIRPOL project was initiated and supported by Nagoya University, National Astronomical Observatory of Japan, and the University of Tokyo in collaboration with the South African Astronomical Observatory under the financial support of Grants-in-Aid for Scientific Research on Priority Area (A) No. 10147207 and No. 10147214, and Grants-in-Aid No. 13573001 and No. 16340061 of the Ministry of Education, Culture, Sports, Science, and Technology of Japan. MT and RK acknowledge support by the Grants-in-Aid (Nos. 16077101, 16077204, 16340061, 21740147, 26800111, and 19K03922).

\clearpage 

\begin{figure}[t]
\begin{center}
 \includegraphics[width=6.5 in]{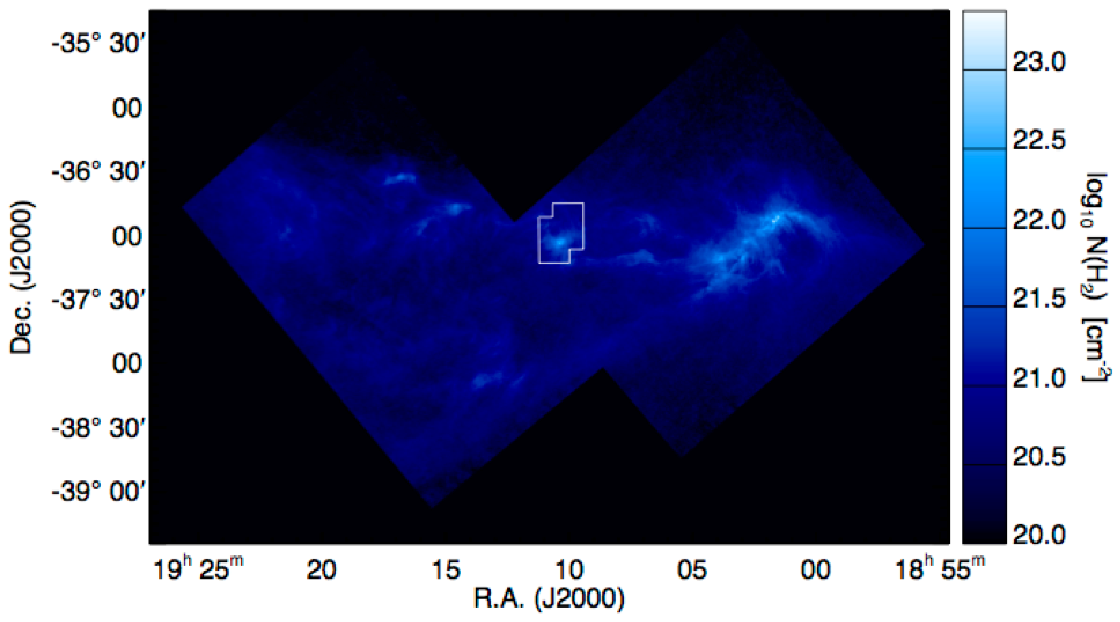}
\end{center}
 \caption{Column density map based on the {\it Herschel} satellite (Andr\'{e} et al. 2010; Bresnahan et al. 2018). The region surveyed with NIR polarimetry is enclosed by the white line.}
   \label{fig}
\end{figure}

\clearpage 

\begin{figure}[t]
\begin{center}
 \includegraphics[width=6.5 in]{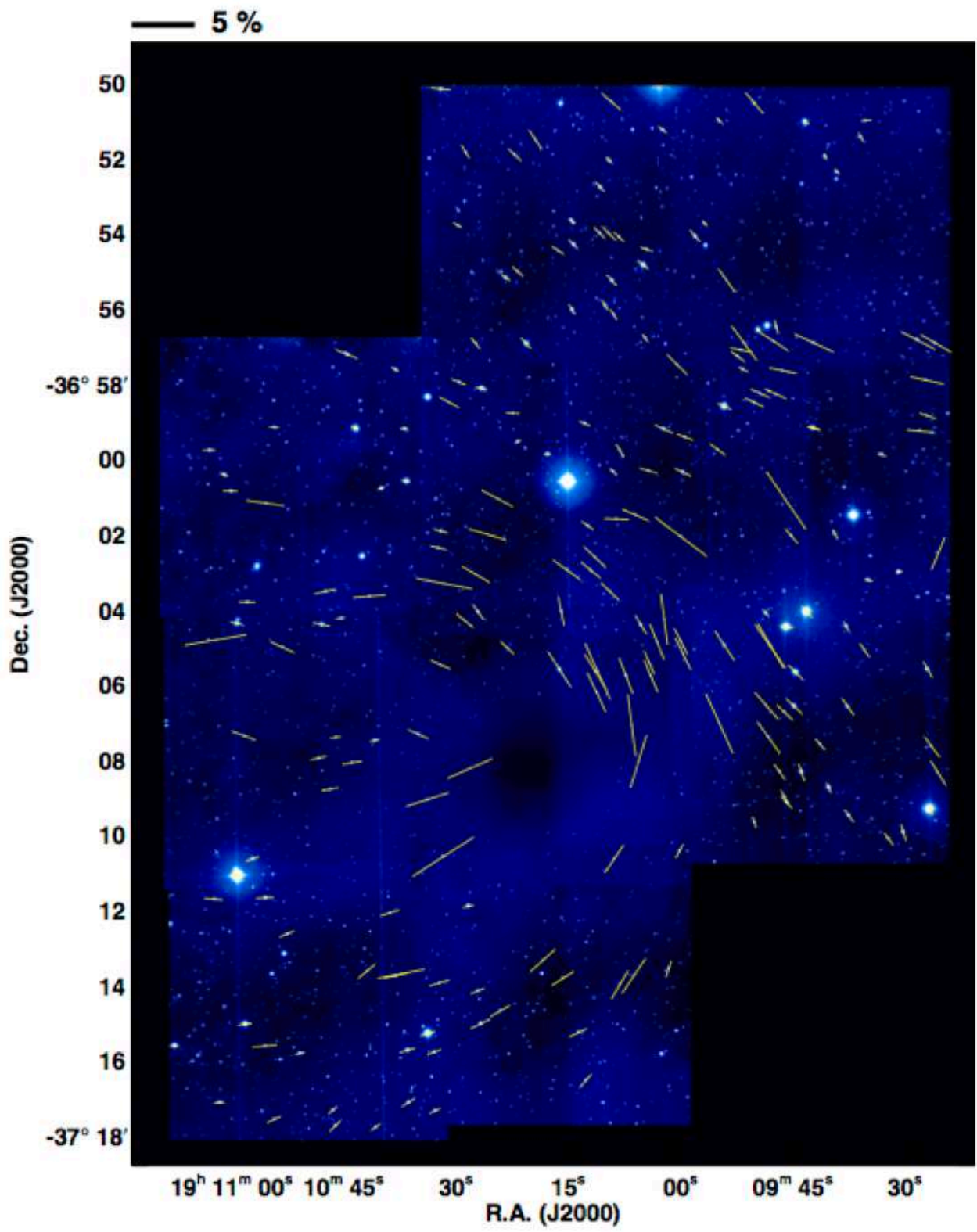}
\end{center}
 \caption{Polarization vectors of point sources superimposed on the mosaic intensity image in the $H$ band. Stars for which $P_H \ge 0.5\%$ and $P_H / \delta P_H \ge 5$ are shown. The scale bar above the image indicates 5\% polarization.}
   \label{fig}
\end{figure}

\clearpage 

\begin{figure}[t]
\begin{center}
 \includegraphics[width=6.5 in]{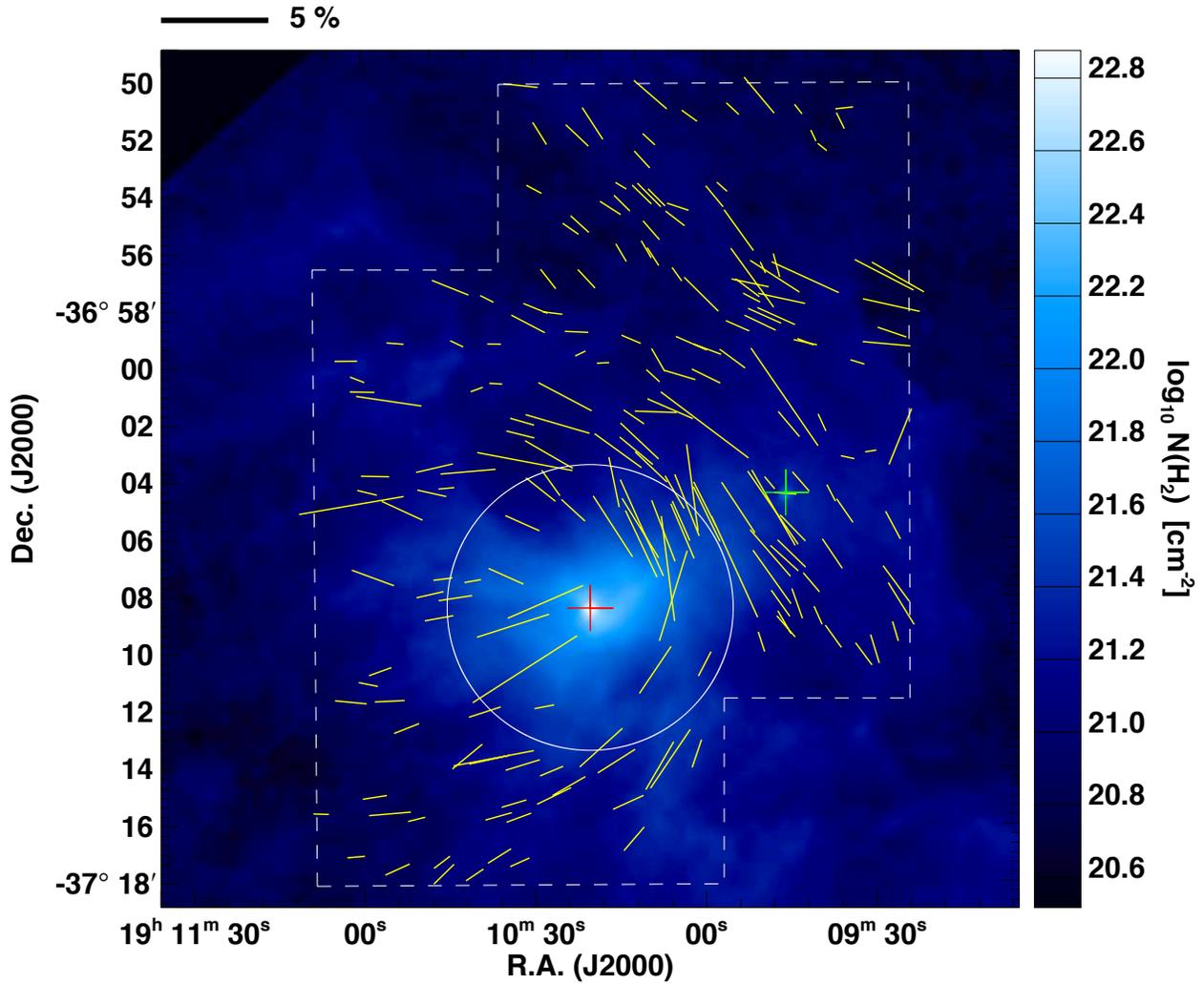}
\end{center}
 \caption{Polarization vectors of point sources in the $H$ band superimposed on the column density map based on {\it Herschel}. The region surveyed with NIR polarimetry is enclosed by the dashed white line. The core radius ($300''$) is indicated by the white circle, and the center of SL42 is shown by the red plus sign. The green plus sign indicates the young star H${\alpha}$16. The scale bar above the image indicates 5\% polarization.}
   \label{fig}
\end{figure}

\clearpage 

\begin{figure}[t]
\begin{center}
 \includegraphics[width=6.5 in]{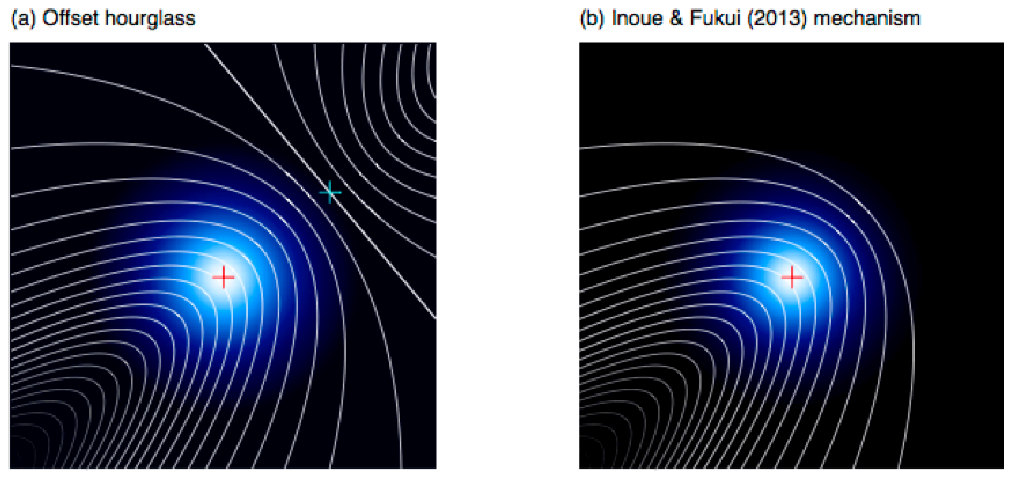}
\end{center}
 \caption{Schematic figures to explain the origin of the curved magnetic fields (white lines) observed toward SL42. (a) Offset hourglass, with an offset angle between the center of mass (red plus sign) and the center of the hourglass-shaped magnetic fields (blue plus sign). The structure can be generated by the accumulation of an initially nonuniform medium dragging magnetic fields. (b) Inoue \& Fukui (2013) mechanism. A shock wave propagates from the upper-right corner to the lower-left corner, sweeping the magnetic fields, and the magnetic fields wrap around the core to create the curved magnetic field structure. In both figures, the darkness of the contour refers to the expected polarization level.}
   \label{fig}
\end{figure}

\clearpage 

\begin{figure}[t]
\begin{center}
 \includegraphics[width=6.5 in]{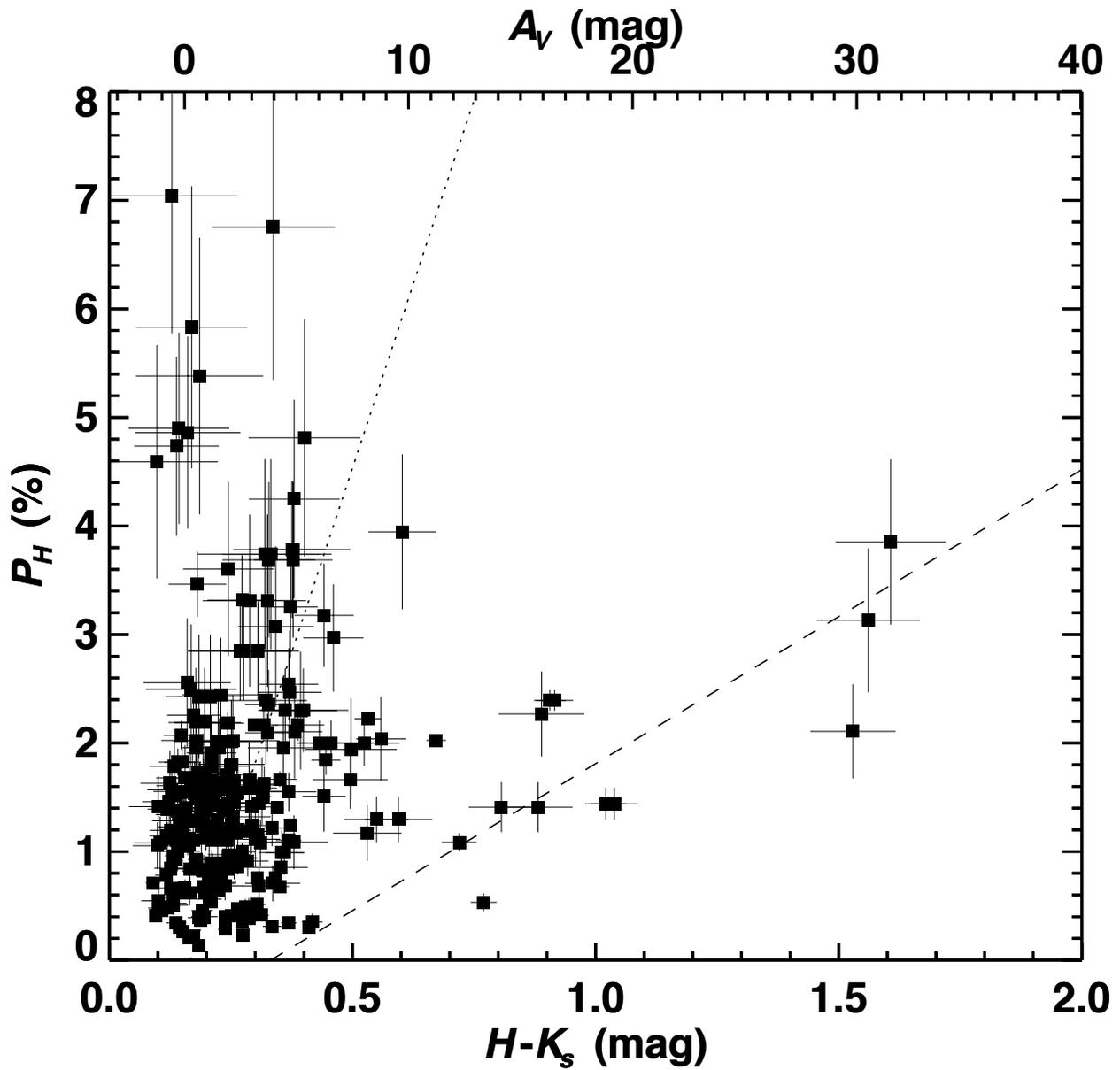}
\end{center}
 \caption{Relationship between the polarization degree and $H-K_{\rm s}$ color toward the background stars. Stars for which $P_H / \delta P_H \ge 4$ are plotted. The dashed line and dotted line denote linear fits to the stars with $H-K_{\rm s} \ge 0.7$ mag and to all the stars.}
   \label{fig}
\end{figure}

\clearpage 

\begin{figure}[t]
\begin{center}
 \includegraphics[width=6.5 in]{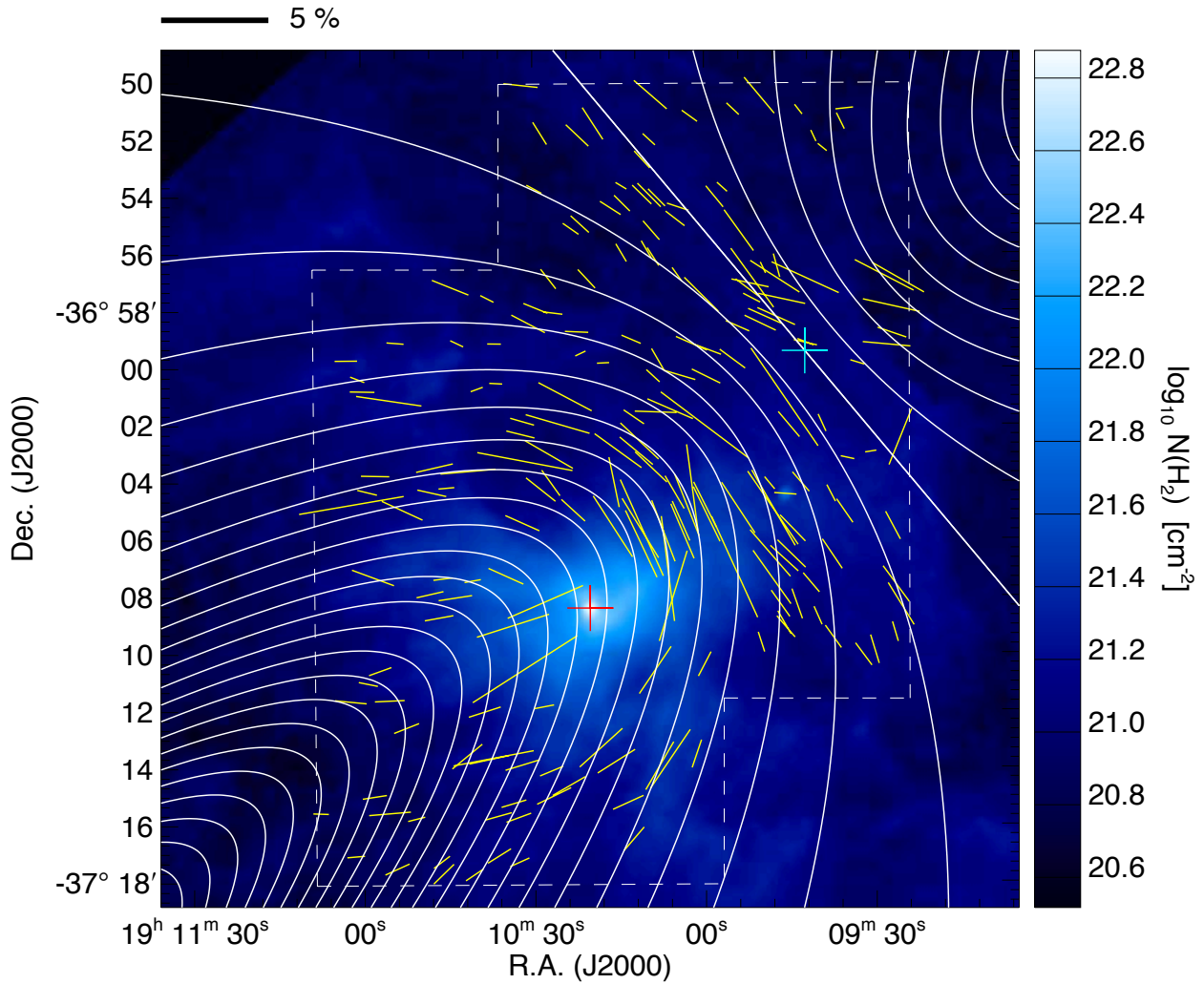}
\end{center}
 \caption{Polarization vectors of point sources in the $H$ band superimposed on the column density map based on {\it Herschel}. The region surveyed with NIR polarimetry is enclosed by the dashed white line. The center of SL42 is shown by the red plus sign. The blue plus sign indicates the center of the hourglass-shaped magnetic fields. The white lines indicate the direction of the magnetic field inferred from the parabolic fitting. The scale bar above the image indicates 5\% polarization.}
   \label{fig}
\end{figure}

\clearpage 

\begin{figure}[t]
\begin{center}
 \includegraphics[width=6.5 in]{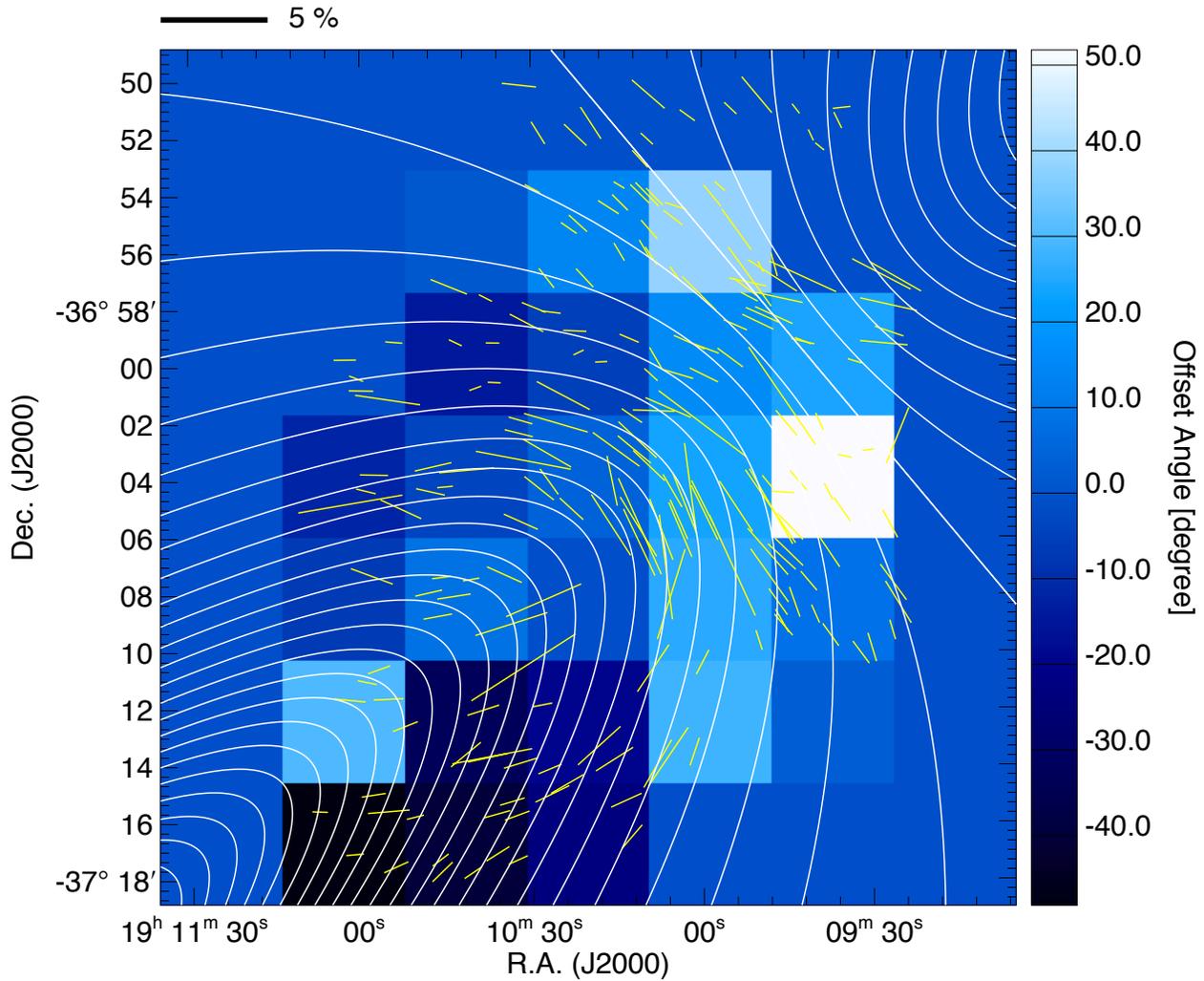}
\end{center}
 \caption{Polarization vectors of point sources in the $H$ band superimposed on the offset angle map. In each grid box with $300''$ width, the average angle differences between the observations and the parabolic model were calculated. The white lines indicate the direction of the magnetic field inferred from the parabolic fitting. The scale bar above the image indicates 5\% polarization.}
   \label{fig}
\end{figure}

\clearpage 

\begin{figure}[t]
\begin{center}
 \includegraphics[width=6.5 in]{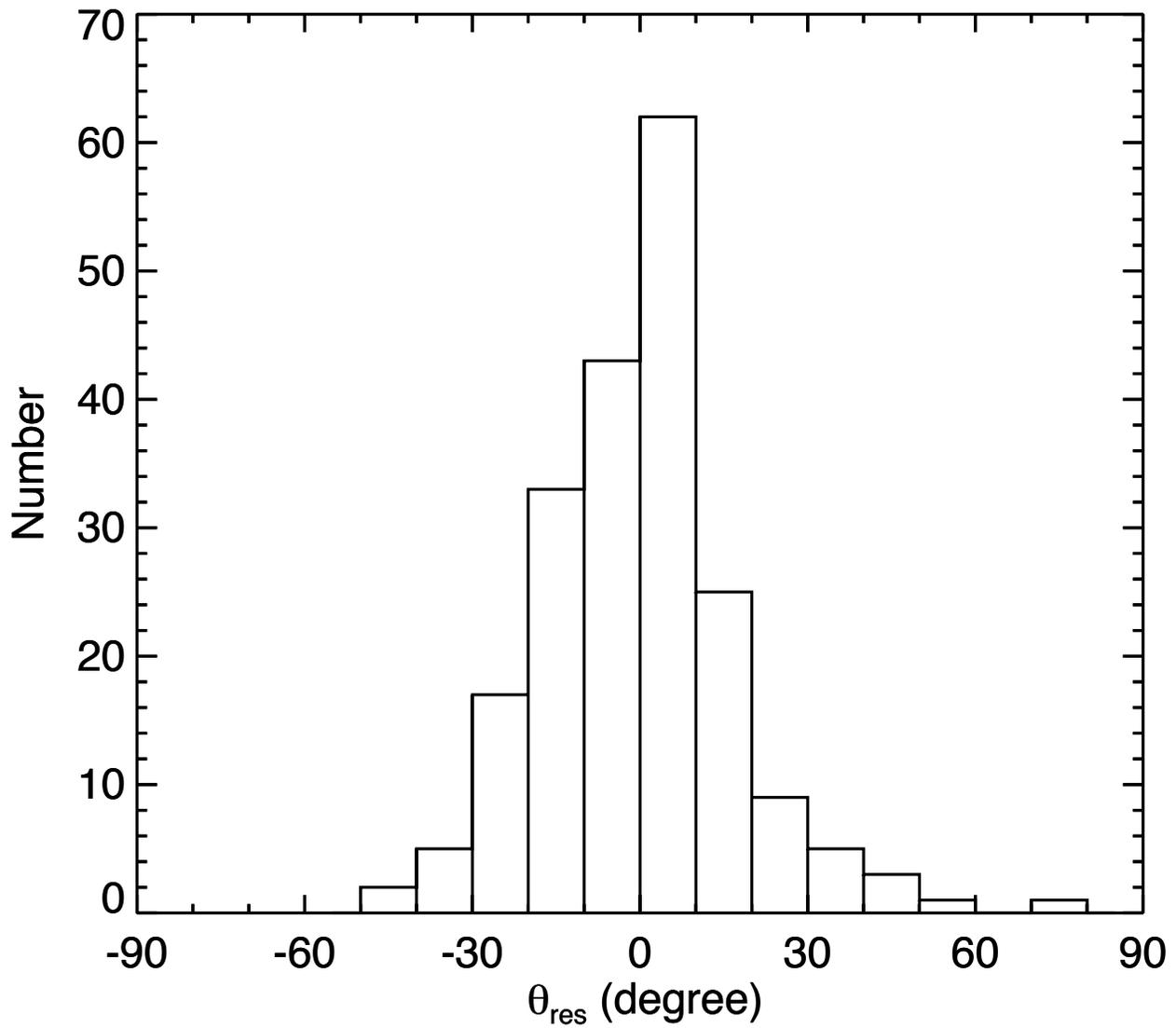}
\end{center}
 \caption{Histogram of the residuals for the observed polarization angles after subtraction of the angles obtained by parabolic fitting ($\theta_{\rm res}$) and its average angle offset measured in a $300''$ width box.}
   \label{fig}
\end{figure}

\clearpage 

\begin{figure}[t]
\begin{center}
 \includegraphics[width=6.5 in]{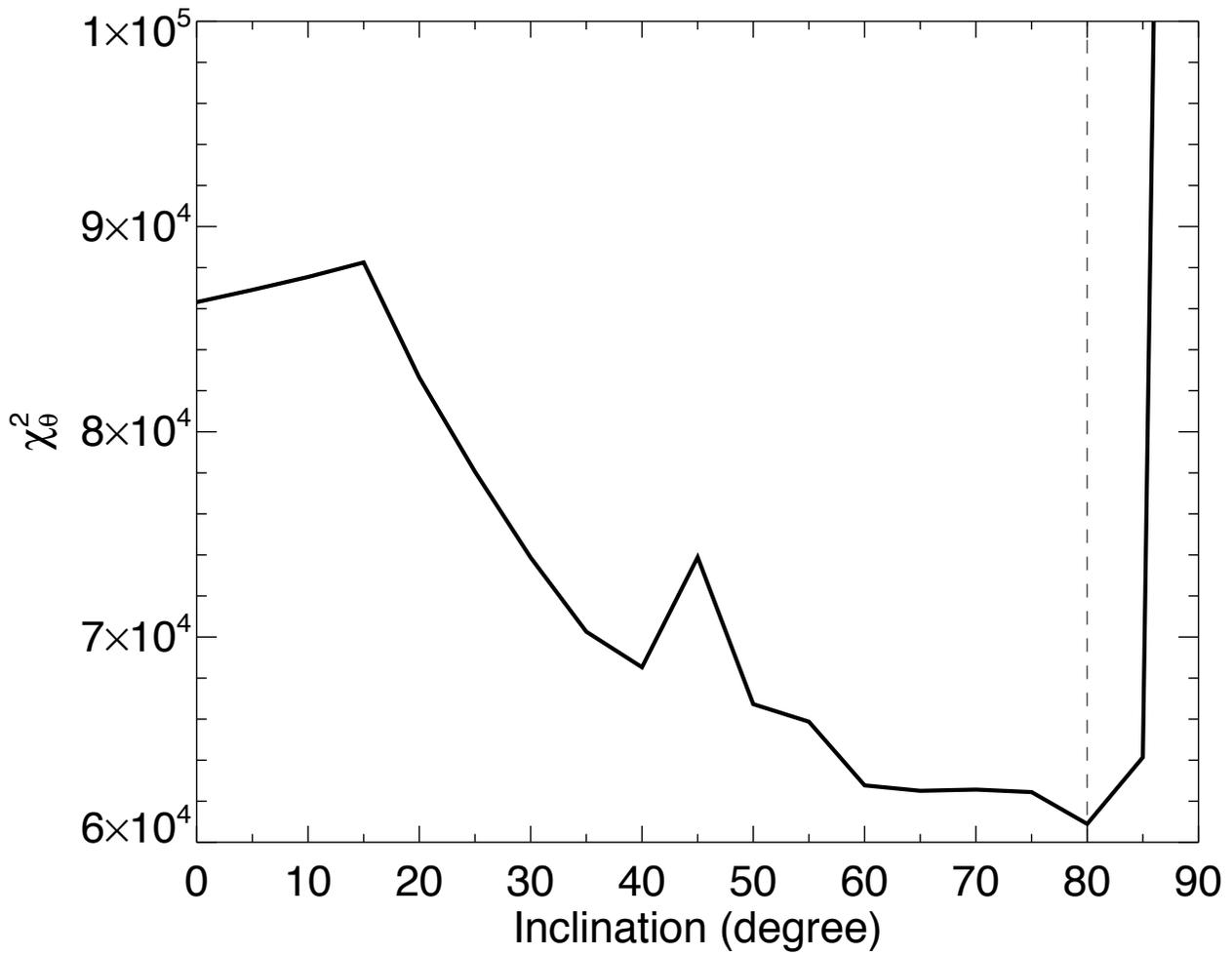}
\end{center}
 \caption{$\chi^2$ distribution of the polarization angles ($\chi_{\theta}^2$). The best magnetic curvature parameter ($C$) was determined at each $\gamma_{\rm mag}$. $\gamma_{\rm mag}=0^{\circ}$ and $90^{\circ}$ corresponds to the edge-on and pole-on geometry in the magnetic axis.}
   \label{fig}
\end{figure}

\clearpage 

\begin{figure}[t]
\begin{center}
 \includegraphics[width=6.5 in]{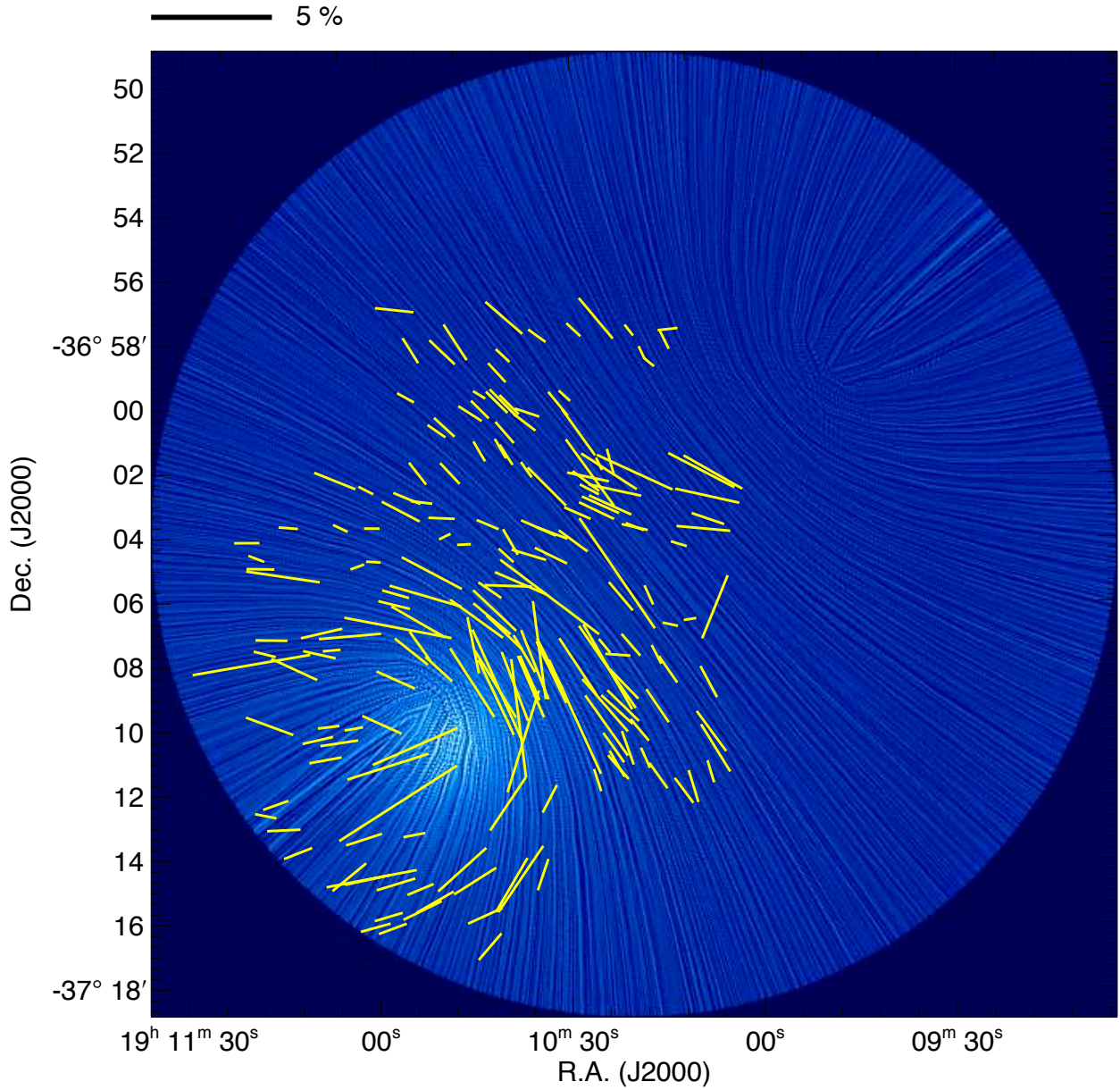}
\end{center}
 \caption{Best-fit 3D parabolic model with observed polarization vectors (yellow vectors). The background image was generated using the line integral convolution (LIC) technique (Cabral \& Leedom 1993). The direction of the LIC \lq \lq texture'' is parallel to the direction of the magnetic fields. The background image is based on the column density of the model core. The scale bar above the image indicates 5\% polarization.}
   \label{fig}
\end{figure}

\clearpage 

\begin{figure}[t]
\begin{center}
 \includegraphics[width=6.5 in]{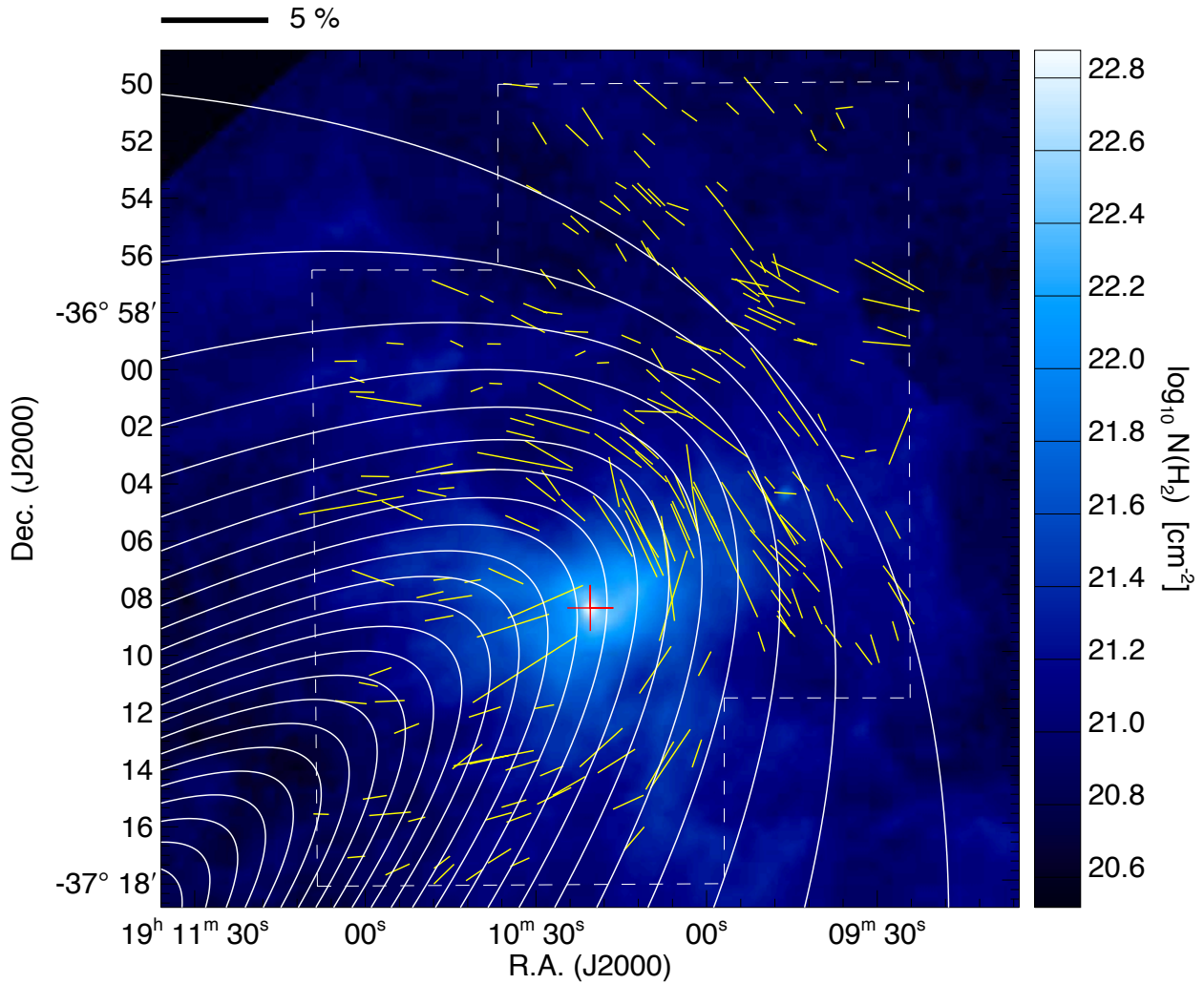}
\end{center}
 \caption{Polarization vectors of point sources in the $H$ band superimposed on the column density map based on {\it Herschel}. The region surveyed with NIR polarimetry is enclosed by the dashed white line. The center of SL42 is shown by the red plus sign. The white lines indicate the direction of the magnetic field inferred from the parabolic fitting. These are rough approximation that take into account the characteristics of the Inoue \& Fukui (2013) model. The scale bar above the image indicates 5\% polarization.}
   \label{fig}
\end{figure}

\clearpage 

\begin{figure}[t]
\begin{center}
 \includegraphics[width=6.5 in]{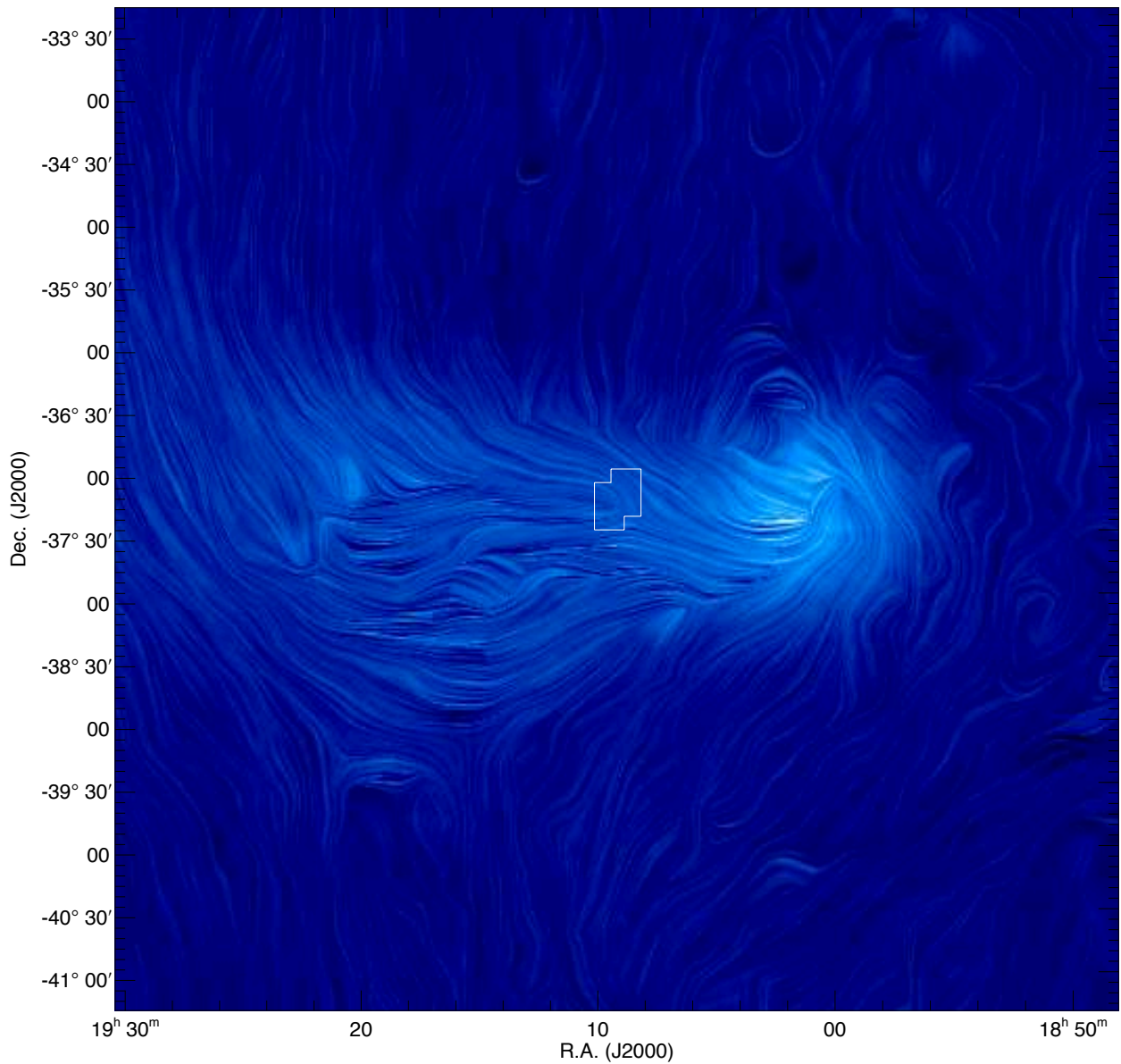}
\end{center}
 \caption{Magnetic field map of the CrA complex. The region surveyed with NIR polarimetry is enclosed by the white line. The background image was generated using the line integral convolution (LIC) technique (Cabral \& Leedom 1993). The direction of the LIC \lq \lq texture'' is parallel to the direction of the magnetic fields. The background image is based on the Stokes $I$ image of the $Planck$ data (353 GHz). The resolution of the image is $5'$.}
   \label{fig}
\end{figure}

\end{document}